\documentclass[prl,aps,twocolumn,preprintnumbers,showpacs,superscriptaddress]{revtex4}
\usepackage{latexsym,epsfig,amssymb,amsfonts,amsmath,graphicx,bbm}

\newcommand{\ket}[1]{\left | \, #1 \right \rangle}

\newcommand{\vac}{\ket{\textrm{vac}}}

\newcommand{\eqr}[1]{Eq.~(\ref{#1})}
\newcommand{\fir}[1]{Fig.~\ref{#1}}

\allowdisplaybreaks

\begin{document}

\title{Efficient generation of graph states for quantum computation}

\author{S.~R. \surname{Clark}}%
\affiliation{Clarendon Laboratory, University of Oxford, Parks Road, Oxford OX1 3PU, U.K.}%
\author{C. \surname{Moura Alves}}%
\affiliation{Clarendon Laboratory, University of Oxford, Parks Road, Oxford OX1 3PU, U.K.}%
\affiliation{DAMTP, University of Cambridge, Wilberforce Road, Cambridge CB3 0WA, U.K.}%
\author{D. \surname{Jaksch}}%
\affiliation{Clarendon Laboratory, University of Oxford, Parks Road, Oxford OX1 3PU, U.K.}%

\date{\today}

\begin{abstract}
We present an entanglement generation scheme which allows
arbitrary graph states to be efficiently created in a linear
quantum register via an auxiliary entangling bus. The dynamics of
the entangling bus is described by an effective non-interacting
fermionic system undergoing mirror-inversion in which qubits,
encoded as local fermionic modes, become entangled purely by Fermi
statistics. We discuss a possible implementation using two species
of neutral atoms stored in an optical lattice and find that the
scheme is realistic in its requirements even in the presence of
noise.
\end{abstract}


\maketitle

Bipartite entanglement has long been recognized as a useful
physical resource for tasks such as quantum cryptography and
quantum teleportation. Similarly, multipartite entanglement is an
essential ingredient for more complex quantum information
processing (QIP) tasks, and interest in this resource has grown
since its controlled generation was demonstrated in several
physical systems~\cite{OLattice,Multi}. An important class of
multipartite entangled states are graph states. By using vertices
in a graph to represent qubits, and edges to represent an Ising
type interaction that has taken place between two qubits, the
graph formalism gives an effective characterization of
entanglement by the presence of edges~\cite{Graph}. Special
instances of graph states are the resource used in multi-party
communication protocols, in quantum error correcting
codes~\cite{QEC} and in one-way quantum computing~\cite{Cluster}.

Initial proposals for the generation of graph states in physical
systems focussed on qubit lattices of fixed geometry, where each
qubit interacts only with its nearest-neighbors~\cite{Graph}. Such
a scheme has been experimentally implemented in 1D with optical
lattices of neutral atoms via controlled
collisions~\cite{OLattice}. The graph states generated with this
method follow the geometry of the lattice, and for 2D/3D square
lattices they constitute, together with single qubit measurements,
a universal resource for quantum computation~\cite{Cluster}.
However, the generation of more complex graphs, where the set of
edges does not translate into a regular arrangement of qubits,
e.g.~the quantum Fourier transform graph state, requires the
ability to pre-engineer a complicated geometry of the qubit
interactions. A simple scheme in which any graph state can be
created in a set of qubits with a regular fixed geometry is
therefore highly desirable, and some progress has been made
towards this with non-deterministic linear optical
protocols~\cite{Optics}.

\begin{figure}
\includegraphics[width=7.5cm]{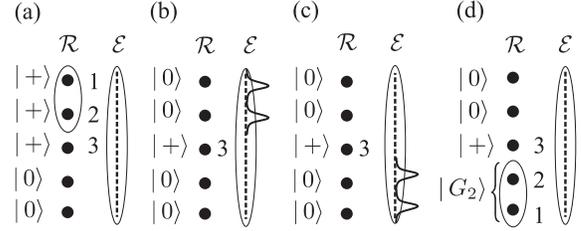}
\caption{(a) Consider a quantum register $\mathcal{R}$ which has 3
graph qubits in a state $\ket{+}$. (b) Two of them are transferred
to the EB $\mathcal{E}$ where their state is mapped into local
fermionic modes. (c) $\mathcal{E}$ evolves via $H_{\textrm{f}}$
for time $\tau$, which results in the mirror-inversion of the two
qubits. (d) The qubits at the mirror-inverted location are
transferred back to $\mathcal{R}$, yielding a graph state with 2
vertices $\ket{G_2}$. Repeating this procedure with different
qubits allows any 3-qubit graph state to be generated in
$\mathcal{R}$.}\label{scheme}
\end{figure}

In this paper we propose a scheme for efficiently generating
arbitrary graph states within a linear quantum register via an
auxiliary entangling bus (EB). The EB is a fixed `always on'
system, equivalent, in a specific limit, to a non-interacting
fermionic system where qubits transferred from the register
subsequently become encoded as local fermionic modes
(LFMs)~\cite{Kitaev}. The dynamics of the EB over a fixed time
then result in the complete mirror-inversion of these LFMs
yielding robust phases within its fermionic state purely as a
consequence of Fermi statistics. Upon transferring the
mirror-inverted LFMs back to the register the EB has generated
entanglement between the qubits. For two qubits this procedure
performs a controlled-$\sigma_z$ (c-$\sigma_z$) gate, in addition
to the inversion~\cite{Bose}, as depicted in~\fir{scheme}, and
this is sufficient to establish an edge between the vertices that
these qubits represent~\cite{Graph}. We examine an implementation
of this scheme where the EB is an XY spin chain formed from an
optical lattice of neutral atoms and investigate numerically its
fidelity in the presence of noise. Finally we describe how,
through the effective quantum circuit implemented by transferring
and evolving more than two qubits within the EB for the same fixed
time, arbitrary graph states of $n$ vertices can be generated in
at most {\em O}$(2n)$ EB cycles. This represents an improvement
over the {\em O}$(n^2)$ steps required in a network model composed
of two-qubit gates.

The starting point for our implementation of the EB is a 1D
optical lattice of ultracold bosonic atoms in two long-lived
internal (hyperfine) states $\ket{a}$ and $\ket{b}$. The dynamics
of this system over $N$ sites is then described by a two-species
Bose-Hubbard model (BHM) given by~\cite{Duan}
\begin{eqnarray}
H &=& \sum_{n=1}^{N}\left(\frac{U^a_n}{2}a^{\dagger2}_n{a_n^2} +
\frac{U^b_n}{2}b^{\dagger2}_n{b_n^2} +
U^{ab}_na^{\dagger}_na_nb^{\dagger}_nb_n\right) \nonumber \\
&&-\sum_{n=1}^{N-1}\left(t^a_na^{\dagger}_na_{n+1}+
t^b_nb^{\dagger}_nb_{n+1}+ \textrm{H.c.}\right) + H_B, \quad
\end{eqnarray}
where $a_n(b_n)$ is the bosonic destruction operator for an
$a(b)$-atom in the $n$th site, and
$H_B=(B/2)\sum_n(a^{\dagger}_na_n-b^{\dagger}_nb_n)$ is the
contribution of a uniform external field. The parameters
$t^{a(b)}$ and $U^{a(b)}$, $U^{ab}$ are the
laser-intensity-dependent hopping matrix elements and on-site
interactions for atoms in states $\ket{a}(\ket{b})$ respectively.
These parameters will in general have a spatial profile across the
lattice. The dynamic controllability and long decoherence times of
this system have made it of considerable interest for
QIP~\cite{Knight,Toolbox} and for realizing spin
models~\cite{Duan}.

We focus on the two-species BHM in the limit of large
interactions, $U^a, U^b, U^{ab} \gg t^a, t^b$, which energetically
prohibit the multiple occupancy of any site. Hopping can be then
treated perturbatively and to lowest order, for an initial Mott
insulating state with commensurate filling of one atom per lattice
site, the effective Hamiltonian is found to be~\cite{Duan}
\begin{equation}
H_{\textrm{s}} = \sum_{n=1}^{N-1} \lambda^{zz}_n
\sigma_n^z\sigma_{n+1}^z + \lambda^{z}_n\sigma_n^z -
\lambda^{xy}_n (\sigma_n^x\sigma_{n+1}^x +
\sigma_n^y\sigma_{n+1}^y).
\end{equation}
Hence we obtain the anisotropic Heisenberg spin model in the
optical lattice, with $\ket{a}\equiv\ket{\uparrow}$ and
$\ket{b}\equiv\ket{\downarrow}$ at each site. The corresponding
Pauli operators are then $\sigma_n^z= a_n^{\dagger} a_n -
b_n^{\dagger} b_n$, $\sigma_n^x= a_n^{\dagger} b_n + b_n^{\dagger}
a_n$, $\sigma_n^y= -i(a_n^{\dagger} b_n - b_n^{\dagger} a_n)$,
while the couplings are given by $\lambda^{zz}_n = (t^{a2}_n +
t^{b2}_n)/(2 U^{ab}_n) - t^{a 2}_n/U^a_n - t^{b2}_n/U^b_n$,
$\lambda^{z}_n = 4(t^{a2}_n/U^a_n - t^{b2}_n/U^b_n)+B/2$, and
$\lambda^{xy}_n = t^a_n t^b_n /U^{ab}_n$. The construction of the
EB requires the optical lattice parameters to be engineered such
that $U^a_n=U^b_n=2U^{ab}_n$ and $t^a_n=t^b_n$, thereby ensuring
that $\lambda^{zz}_n=0$, $\lambda^{z}_n=B/2$, and that
$H_{\textrm{s}}$ reduces to a pure XY spin chain. For simplicity
we assume that only the hopping possesses spatial dependence via
$t_n^a=t_n^a=T\sqrt{\alpha_n}$, with a profile obeying $0 <
\alpha_n \leq 1$ so $\max(t_n^{a(b)})\leq T$, and that the
interaction energies are constant over the system as
$U^a=U^b=2U^{ab}=U$ with both $U$ and $T$ constants. To first
order in $T/U$ the dynamics of the optical lattice reduces to an
XY spin chain with couplings $\lambda^{xy}_n=(2T^2/U)\alpha_n$.
For the moment we shall assume ideality where $U/T \gg 1$;
however, later we will investigate how the fidelity of this
mapping varies with $U/T$ and in the presence of noise.

It is well known that the Jordan Wigner transformation
(JWT)~\cite{Sachdev} maps the XY spin chain to a non-interacting
fermionic Hamiltonian
$H_{\textrm{f}}=-\sum_nj_n(c_n^{\dagger}c_{n+1} +
c_{n+1}^{\dagger}c_n)+\sum_nu_nc_n^{\dagger}c_n$, where $j_n =
2\lambda^{xy}_n$, $u_n = B$, and $c_n$ is a fermionic destruction
operator for the $n$th site obeying the usual anticommutation
relations. We are particularly interested in the {\em angular
momentum} hopping profile~\cite{StateTransfer} given by $j_n =
(J/2)\sqrt{n(N-n)}$, which we write as $j_n = W\alpha_n$ with
$\alpha_n = 2\sqrt{(n/N)[1-(n/N)]}$ and $W=4T^2/U=JN/4$ so
$\max(j_n)\leq W$. In this case the projection of $H_{\textrm{f}}$
onto the single fermion subspace of the lattice, $\mathcal{H}_1$,
results in a Hamiltonian equivalent to $H_1 = -JS_x +
B\mathbbm{1}$, where $S_x$ is the $x$ angular momentum operator
for an `effective' spin-$\mathcal{S}$ particle, with
$\mathcal{S}=(N-1)/2$. The single-fermion states
$\{\ket{n}=c^{\dagger}_n\vac\}$ then correspond to the $z$-angular
momentum eigenstates $\{\ket{\mathcal{S},l}_z\}$ of the
spin-$\mathcal{S}$ particle, with $\ket{1} =
\ket{\mathcal{S},-\mathcal{S}}_z, \dots, \ket{N} =
\ket{\mathcal{S},\mathcal{S}}_z$. The dynamics generated in
$\mathcal{H}_1$, when $H_1$ is applied for a fixed time $\tau =
\pi/J$, result in the time-evolution unitary $U_1(\tau) =
\exp{(i\phi_B)}\exp{(i\pi S_x)}$ composed of an overall phase
$\phi_B = -B\pi/J$ for $\mathcal{H}_1$ and a rotation of the
spin-$\mathcal{S}$ particle about the $x$-axis by $\pi$. This
leads directly to perfect state transfer over the
lattice~\cite{StateTransfer}. The action of $U_1(\tau)$ on the
single-particle basis follows from its equivalence to the
$z$-angular momentum states where $\exp{(i\pi
S_x)}\ket{\mathcal{S},l}_z =
\exp{(i\pi\mathcal{S})}\ket{\mathcal{S},-l}_z$. Thus we find that
$U_1(\tau)\ket{n}=\exp{(i\phi_1)}\ket{\bar{n}}$, with the phase
$\phi_1 = \pi\mathcal{S} + \phi_B$ and mirror-conjugate location
$\bar{n} = N - n + 1$. Choosing $B=\mathcal{S}J$ is sufficient to
ensure that the single-particle phase $\phi_1$ vanishes. The
evolution of the fermionic modes $c^{\dagger}_n$ then satisfies
$Uc^{\dagger}_n U^{\dagger} = c^{\dagger}_{\bar{n}}$, where $U =
\exp{(-iH_{\textrm{f}}\tau)}$, and the dynamics of the system
describe the complete mirror-inversion of the LFMs.

Under the JWT the $N$ qubit (or spin) states
$\ket{q_1,\dots,q_N}$, with $q_n \in \{0,1\}$, of the chain are
mapped to Fock states of the LFMs as $\ket{q_1,\dots,q_N} =
(c^{\dagger}_1)^{q_1}\dots(c^{\dagger}_N)^{q_N}\vac$ (using
lattice operator ordering) which describe the occupancy of the
system by quasi-fermions. The use of fermionic mode occupancy as a
basis for quantum computation has been proposed
before~\cite{Kitaev}. Complications arise since the accessibility
of the Fock space of real fermions is restricted by {\em
superselection} rules, and also, in contrast to bosons, it does
not possess a natural tensor product structure permitting
independent operations on each mode, thus making it intrinsically
nonlocal~\cite{ZanardiWu}. However, since we have focussed on a
physical system which {\em maps} to a non-interacting fermionic
system, the Fock space is fully accessible, thus enabling
superpositions of states with different numbers of quasi-fermions,
which are essential for encoding qubits.

For systems of identical fermions a bilinear fermionic Hamiltonian
such as $H_{\textrm{f}}$ suffices to generate mode entanglement,
despite describing a non-interacting system~\cite{Vedral}. The
entanglement generated by mirror-inversion then follows
straightforwardly from Fermi statistics through its action on Fock
states as
\begin{equation}
e^{-iH_{\textrm{f}}\tau}\ket{q_1,\dots,q_N}=e^{-i\pi\Sigma_Q}\ket{q_N,\dots,q_1},
\label{fock}
\end{equation}
where $\Sigma_Q$ is the number of anti-commutations of the
operators $c^{\dagger}_n$ required to reestablish a Fock state.
Specifically $\Sigma_Q = Q(Q-1)/2$, where $Q$ is the number of
fermions, i.e. $Q =
\sum_{n=1}^{N}c^{\dagger}_nc_n=\sum_{n=1}^Nq_n$, and so phases are
only acquired between subspaces with different $Q$. Since
\eqr{fock} is written in terms of Fock states, the inverse-JWT
removes the antisymmetry when mapping back to qubits, while
leaving the phases acquired between fermion-number (or total
magnetization) subspaces untouched. Thus the evolution of the EB
over a fixed time $\tau$ is equivalent to a quantum circuit
$\mathcal{C}(N)$ composed of c-$\sigma_z$ gates between all
distinct pairs of $N$ qubits followed by the inversion operator
$R$, as shown in \fir{circuitgraph}(a) for $N=5$. Usefully, if
$N-q$ qubits are in the state $\ket{0}$, then this circuit reduces
to $\mathcal{C}(q)$ between the remaining $q$ qubits, independent
of their locations, followed by the full inversion $R$.

\begin{figure}
\includegraphics[width=7.5cm]{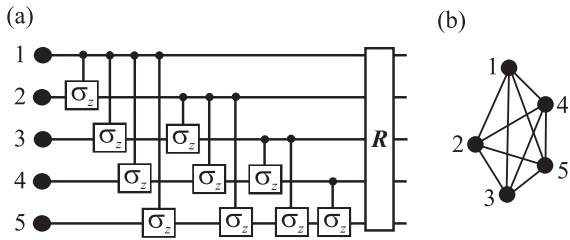}
\caption{For 5 qubits we have (a) the quantum circuit
$\mathcal{C}(5)$ equivalent to the dynamics of $H_{\textrm{f}}$
for a time $\tau$, and (b) the fully connected graph state
$\ket{G_5}$ with 5 vertices generated by this circuit if all the
qubits are initialized in the $\ket{+}$ state. Both the circuit
and the resulting graph state generalize in an obvious way for
more qubits.}\label{circuitgraph}
\end{figure}

In passing we note that the above results apply to more general
settings. Suppose we partition the fermionic lattice into $M$
equal blocks, each labelled by their central site $k$, and
composed of sites $m(k)$. Within each block $k$ we consider an
extended fermionic mode $f_k^{\dagger} = \sum_{n \in m(k)}
\phi^k_n c_n^{\dagger}$, defined by a single-particle state
$\phi^k_n$ contained entirely within the block $k$ and symmetrical
about its center. The dynamics of the EB over some fixed time will
be equivalent to $\mathcal{C}(M)~R$ as long as $j_n$ and $u_n$ are
chosen such that the dynamics of $H_{\textrm{f}}$ performs
mirror-inversion with respect to $f_k^{\dagger}$. We can equally
consider partitioning a 1D continuous fermionic system, described
by field operators $\hat{\psi}^{\dagger}(x)$, and defining
extended LFMs analogously as $f^{\dagger}_k = \int_{m(k)}
dx~\phi^k(x) \hat{\psi}^{\dagger}(x)$, with $\phi^k(x)=0, ~
\forall~x \notin m(k)$. In this case, harmonic trapping
$V(x)=m\omega^2x^2/2$ and Gaussian modes $f^{\dagger}_k$ are
sufficient for mirror-inversion over a time $\tau=\pi/\omega$,
giving an arrangement conceptually similar to the 1D
cold-collision proposal in~\cite{Calarco} at the Tonks-Girardeau
limit~\cite{Paredes}.

The generation of arbitrary graph states requires the EB to be
augmented with a linear register of $N$ qubits, and a transfer
process which maps a qubit state into a LFM in the EB via
$\sigma_n^{+} \mapsto c^{\dagger}_n$, where $\sigma_n^{+}$ is the
Pauli ladder operator. The register and the EB are then
initialized in the states $\otimes_i^N\ket{0}_i$ and $\vac$
respectively. The scheme begins by choosing a set of register
qubits $\Gamma$ to be the graph vertices, and applying a Hadamard
transformation to each of them: $\ket{0} \rightarrow \ket{+}=
(\ket{0} + \ket{1})/\sqrt{2}$, as in~\fir{scheme}(a) for qubits
$1-3$. A subset $\Sigma$ of $m$ of these qubits is then
transferred to the EB and allowed to evolve for a time $\tau$, as
shown (for $m=2$) in~\fir{scheme}(b) and~\fir{scheme}(c) for
qubits $1$ and $2$. The LFM qubits within the EB at the
corresponding mirror-inverted locations $\bar{\Sigma}$ are then
transferred back to the register, yielding a fully connected graph
state $\ket{G_m}$ between these $m$ vertices, as
in~\fir{scheme}(d). Such a state is locally equivalent to a
$m$-qubit GHZ state, as depicted in~\fir{circuitgraph}(b) for
$m=5$. Overlap between EB and register graph qubits can be avoided
by choosing $|\Gamma| = \lceil{N/2}\rceil$ with locations in the
first half of the register.

Our system can generate any graph state of $n$ vertices in at most
{\em O}$(n^2)$ steps by utilizing only the two-qubit interaction
of the EB to establish each edge individually, mimicking the
network model of two-qubit gates. However, by exploiting the
multi-qubit circuit implemented by the EB dynamics over the same
time $\tau$, as in~\fir{circuitgraph}(a), our scheme can improve
this upper bound. Specifically, we proceed iteratively with $g=1$
by (i) transferring the $g$th graph qubit, and all graph qubits
$g_c > g$ which will connect to $g$, into the EB; (ii) allowing
them to evolve for a time $\tau$ creating a complete set of
connections between these vertices, c.f.~\fir{circuitgraph}(b);
(iii) then transferring qubit $g$ back to the register while
leaving the qubits $g_c$ to evolve for one cycle longer in the EB,
subsequently removing all the connections between them; (iv) and
finally transferring the qubits $g_c$  back to the register and
repeating step (i) with $g\mapsto g+1$. Thus, any graph of $g=n$
vertices can be generated in at most {\em O}$(2n)$ uses of the EB.

In the optical lattice implementation, both the register and the
EB are effective spin chains. The register is created by turning
off the hopping completely~\cite{Toolbox} and is separated from
the EB by a potential barrier (see~\cite{Carolina} for details of
this setup). The transfer process between the register and EB can
then be accomplished either directly via state-independent
tunnelling~\cite{OL,Carolina}, or through the implementation of a
swap gate between the register and EB qubits~\cite{Knight}, on a
time scale much faster than $\tau$.

\begin{figure}
\includegraphics[width=7.5cm]{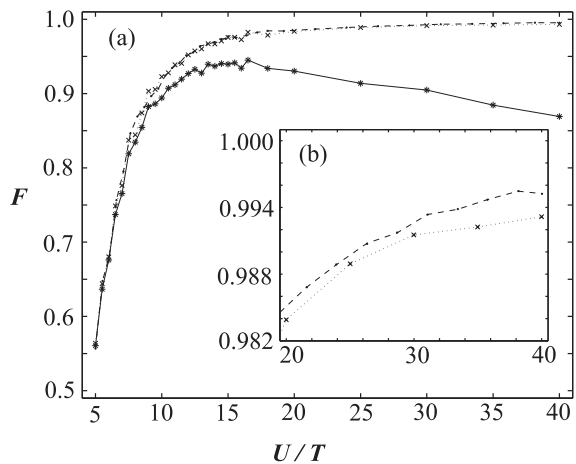}
\caption{(a) The fidelity $F$ of the effective XY spin-chain
implemented by the 2-species BHM with the ratio $U/T$, for no
noise $\Delta=0$ ($\cdot$), $\Delta=1\%$ ($\times$) and
$\Delta=5\%$ ($*$). (b) A close-up of (a). The solid, dashed and
dotted lines are drawn to guide the eye.}\label{plots}
\end{figure}

We considered some dominant sources of imperfections within the
optical lattice EB. In particular, we investigated the fidelity of
the two-species BHM to spin-chain mapping introduced earlier, for
finite $U/T$. We considered a system of size $N=6$ initialized in
a product state $\ket{+} \otimes \ket{0}^{\otimes 4}\otimes
\ket{+}$, and computed the exact time-evolution of the two-species
BHM using the time-evolving block decimation (TEBD)
algorithm~\cite{Vidal} for varying $U/T$ over the appropriate
inversion time $\tau$. Using the effective two-spin density matrix
for the end sites, the fidelity $F$ was computed with the state
$\ket{G_2}$ obtained from a perfectly implemented XY chain. The
simulation results in~\fir{plots}(a) demonstrate that, as
expected, the fidelity increases with $U/T$. Given that
$\tau=UN\pi/16T^2$, increasing the fidelity, by deepening the
lattice, comes at the cost of longer inversion times.
However,~\fir{plots}(b) shows that $F > 0.99$ even at a moderate
ratio of $U/T=26$. At this depth a $\lambda=514$~nm optical
lattice of $^{23}$Na atoms has $\tau=9.3$~ms, while a
$\lambda=826$~nm optical lattice of $^{87}$Rb atoms has
$\tau=79$~ms. These are fast enough for multiple EB inversions to
occur within the decoherence time of the system, which is
typically of order of a second~\cite{OL}.

Finally, we investigated the effect of jitter in the lattice laser
intensities. For $^{87}$Rb the laser intensities $I_a$ and $I_b$
of the $a$ and $b$ lattices were taken as varying independently
according to some Wiener noise $dW_{a(b)}(t)$ with variance
$\Delta^2$: $I_{a(b)}(t) = I_0 + dW_{a(b)}(t)$, where $I_0$ is the
ideal laser intensity. Such laser fluctuations are then
non-linearly related to the corresponding fluctuations in the
hopping $t^{a(b)}_n$ and on-site interaction $U^{a(b)},U^{ab}$
matrix elements of the 2-species BHM~\cite{OL}. We assumed a
simplified version of this noise in which the fluctuations alter
the hopping and interaction scalings $T$ and $U$ contained within
the overall scaling $W$ of $j_n$, but not the spatial profile
$\alpha_n$. Despite this restriction, this noise causes
fluctuations in the inversion time $\tau$ during the dynamics, and
also breaks the symmetry required to ensure that no
$\sigma_n^z\sigma_{n+1}^z$ or spatially-varying $\sigma_n^z$
contributions occur. In~\fir{plots}(a) the fidelity curves are
plotted for $\Delta=1\%$ and $\Delta=5\%$ of $I_0$. For
$\Delta=5\%$ the fidelity is seen to drop off in deeper lattices
due to the cumulative effect of noise over longer inversion times.
Crucially, the fidelity curve suffers only a minimal reduction due
to $\Delta=1\%$ noise, as in~\fir{plots}(b), and this represents a
realistic value for the experimental stabilization of the laser
intensity.

We have shown how arbitrary Graph states can be generated
efficiently by using an EB whose dynamics correspond to a
non-interacting fermionic system undergoing mirror-inversion. By
utilizing an EB which is fixed and always on the dynamical control
required for QIP tasks can be reduced to single qubit operations.
Here an implementation of this scheme using an optical lattice of
neutral atoms was considered in detail. We also note that the EB
properties are well suited to solid state systems. The fidelity of
the optical lattice proposal was examined not only for the depth
ratio $U/T$, but also in the presence of noise, and found to be
both realistic and robust.

This work was supported by the EPSRC IRC network on Quantum
Information Processing (U.K.). S.C. and D.J. thank Peter Zoller
and Hans Briegel for stimulating discussions. C.M.A. thanks Marc
Hein for insightful discussions on graph states and is supported
by the Funda{\c c}{\~a}o para a Ci{\^e}ncia e Tecnologia
(Portugal).

\end{document}